\documentclass[twocolumn,preprintnumbers,aps,prl,floatfix,showpacs,superscriptaddress]{revtex4}
\usepackage{graphicx}
\usepackage{bm}
\usepackage{latexsym}

\begin{document}

\title{Anomaly in Nonlinear Magnetoelectric Response of YbMnO$_{3}$}
\author{U. Adem}
\affiliation{Zernike Institute for Advanced Materials, University of Groningen, 9747
AG Groningen, The Netherlands}
\author{M. Mostovoy}
\affiliation{Zernike Institute for Advanced Materials, University of Groningen, 9747 AG Groningen, The
Netherlands}
\author{N. Bellido}
\affiliation{Laboratoire CRISMAT, UMR CNRS ENSICAEN, 1450 Caen,
France}
\author{A. A. Nugroho}
\affiliation{Zernike Institute for Advanced Materials, University of Groningen, 9747
AG Groningen, The Netherlands} \affiliation{Faculty of Mathematics
and Natural Sciences, Institut Teknologi Bandung, Jl. Ganesha 10,
Bandung 40132, Indonesia}
\author{C. Simon}
\affiliation{Laboratoire CRISMAT, UMR CNRS ENSICAEN, 1450 Caen,
France}
\author{T. T. M. Palstra}
\affiliation{Zernike Institute for Advanced Materials, University of Groningen, 9747
AG Groningen, The Netherlands} \email[Corresponding author:
]{t.t.m.palstra@rug.nl}
\date{\today}

\begin{abstract}
We observe a seemingly complex magnetic field dependence of dielectric constant of hexagonal YbMnO$_3$ near the spin ordering temperature. After rescaling, the data taken at different temperatures and magnetic fields collapse on a single curve describing the sharp anomaly in nonlinear magnetoelectric response at the magnetic transition. We show that this anomaly is a result of the competition between two magnetic phases. The scaling and the shape of the anomaly are explained using the phenomenological Landau description of the competing phases in hexagonal manganites.
\end{abstract}

\maketitle









\newpage

The recent interest in multiferroic materials was triggered by the
discovery of the giant magnetocapacitance (MC) and magnetically-induced
rotations of electric polarization in orthorhombic rare earth
manganites \cite{KimuraNature2003,Goto,Hur}. This multiferroic
behavior is rooted in magnetic frustration, which gives rise to
non-centrosymmetric spin orderings that induce electric polarization
\cite{Cheong}. Furthermore,  the presence of competing spin states
in these frustrated magnets results in a strong sensitivity of the
magnetically-induced electric polarization to applied magnetic
fields. In this respect multiferroics are similar to colossal
magetoresistance manganites and high-temperature superconductors \cite{Dagotto}.


In this Letter we study effects of critical magnetic fluctuations and the competition between different magnetic states on the non-linear magnetoelectric response of the hexagonal YbMnO$_{3}$ by measuring the magnetic field and temperature dependence of its dielectric constant. Ferroelectricity in hexagonal manganites $R$MnO$_3$ ($R$ = Ho-Lu, Y) appears well above the magnetic transition and is of nonmagnetic origin: An electric dipole moment along the $c$ axis is spontaneously induced by tilts of manganese-oxygen bipyramids and buckling of rare earth-oxygen planes at $T_{C} > 600$K \cite{Katsufuji,VanAken,Fennie,Adem}, while the ordering of Mn spins occurs at a much lower temperature $T_{N} < 120$K. However, the spin ordering in hexagonal manganites results in a surprisingly strong lattice relaxation, which affects the spontaneous electric polarization \cite{SeongsuLee}.

The Mn ions in hexagonal manganites form well-separated triangular
layers parallel to the $ab$ plane with antiferromagnetic exchange
interactions between nearest-neighbor spins \cite{Bertaut}, which
makes the Mn spin subsystem low-dimensional and frustrated and
results in enhanced spin fluctuations observed well above $T_{N}$
\cite{JPark}. Frustration and rare earth magnetism are responsible
for a rich variety of magnetic phases observed at low temperatures
and in applied magnetic fields \cite{Lorenz}. Due to magnetoelectric
coupling each magnetic transition gives rise to a singularity of the
dielectric constant \cite{Smolenskii,Huang,Katsufuji,Sugie},
which is more pronounced than the corresponding singularity in
magnetic susceptibility.

We find that close to the N\'{e}el temperature T$_{N}$=81K the MC of YbMnO$_{3}$ measured as a function of magnetic field and temperature obeys a scaling behavior and has a very sharp anomaly. The detailed comparison with results of
model calculations led us to a conclusion that the effect of
magnetic fluctuations is completely overshadowed by the magnetic
field dependence originating from the competition between two
antiferromagnetic states, one of which is weakly ferromagnetic.
Using a mean field Landau expansion of free energy in powers of two
competing order parameters, we reproduce the shape of the anomaly as
well as the changes in the behavior of MC observed
in the wide range of magnetic fields and temperatures.


Polycrystalline samples of YbMnO$_{3}$
were prepared by solid state synthesis. A single crystal was
grown from this powder by the floating zone technique.
Magnetization $M(T)$ of the samples was measured by a Squid
magnetometer (MPMS7 Quantum Design) using a field of 0.5T.
Field dependence of the magnetization was measured up to 5T.
Capacitance of the samples was measured in an commercial system
(PPMS Quantum Design) using a home-made insert and a
Andeen-Hagerling 2500A capacitance bridge operating at a fixed
measurement frequency of 1 kHz as well as using an Agilent 4284A LCR
meter up to 1MHz. Electrical contacts were made
using Ag epoxy.

\begin{figure}[tbp]
\centering
\includegraphics[width=0.4\textwidth]{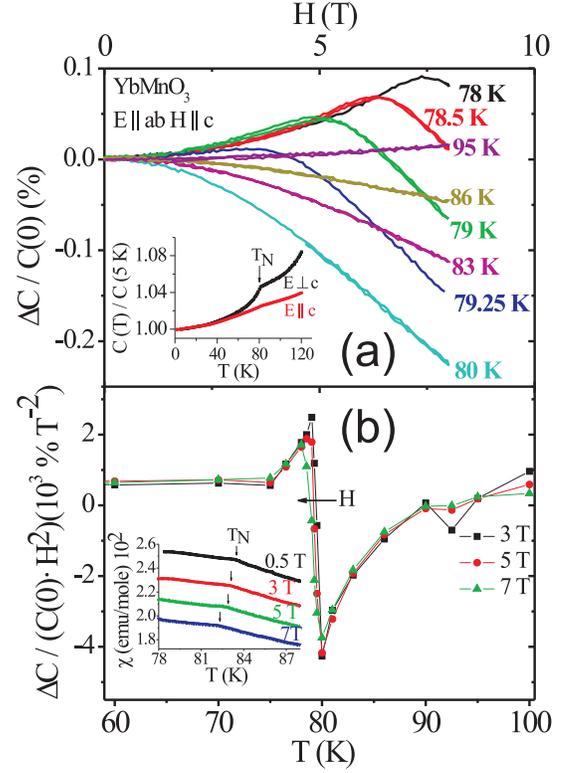}
\caption{(Color online) (a) Magnetic field dependence of MC of YbMnO$_{3}$ single crystal at constant temperatures near T$_{N}$. The temperature dependence of capacitance is added as an inset. (b) Temperature dependence of MC at constant magnetic fields.
The inset shows the shift of T$_{N}$ in magnetic field obtained from magnetic susceptibility measurements. Electric field is parallel to the $ab$ plane, while $H \| c$.} \label{fig:magcap1}
\end{figure}

The temperature dependence of the capacitance $C(T)$ proportional to the in-plane dielectric constant $\varepsilon_{a}$ is shown as an inset of Fig.~\ref{fig:magcap1}a. Below the N\'eel temperature $T_{N} = 82$K, the capacitance is somewhat suppressed by the emergence of magnetic order \cite{Smolenskii}. The MC, $\frac{C(H)-C(0)}{C(0)}$, where $H$ is magnetic field along the $c$ axis, for a set of temperatures between 76.5K and 95K is shown in Fig.~\ref{fig:magcap1}a. In this small temperature interval around T$_{N}$ the behavior changes dramatically: at 76.5K only a positive curvature is observed. With increasing temperature a high-field downturn appears, and at 80K only a negative curvature can be observed, which changes back to positive above 90K.

This unusual behavior is a consequence of the fact that by varying temperature and magnetic field we force the system to pass through a magnetic transition. The critical behavior becomes apparent when we plot $\frac{C(H)-C(0)}{C(0)H^2}$ vs temperature [see Fig.~\ref{fig:magcap1}(b)]. The procedure to evaluate $\Delta$C at fixed magnetic fields versus temperature i.e. replotting the rescaled changes of dielectric constant in magnetic field versus temperature, effectively reveals the magnetic field dependence of C. The strong temperature dependence of C masks the magnetic field dependence when C(T)  is measured at fixed magnetic fields. The data taken at various $T$ and $H$ remarkably fall onto a single curve with a very sharp anomaly at T$_{N}$ where the temperature derivative of MC becomes large and positive while its magnitude shows an almost discontinuous jump from a positive to a negative value.
The observed scaling behavior of MC can be understood in terms of the anomalous nonlinear magnetoelectric response, described by the term $\kappa(T) \left(E_{a}^2 + E_{b}^2\right) H_{c}^2$ in free energy, where $\kappa(T)$ has a singularity at the magnetic transition temperature. The scaling implies that the dependence of magnetic susceptibility for $H\|c$ on electric field $E \| a$ has the same anomaly at T$_{N}$.

The shape of the MC anomaly in YbMnO$_3$ is unusual. The nonlinear magnetoelectric response in antiferromagnets usually originates from the ubiquitous fourth-order coupling of the electric
polarization $P$ to the magnetic order parameter $L$,
$
f_{\rm me} = \frac{g}{2}P^{2}L^{2}.
$
It results in a correction to the bare dielectric
susceptibility $\chi_{0}$,
$
\delta \chi = - g \chi_{0}^{2} \left \langle L
\right \rangle^{2} \propto (- \tau)^{2\beta},$
for $\tau =\frac{T-T_{N}}{T_{N}} < 0,$
which accounts for the observed dielectric constant anomaly below $T_N$ [see the inset of Fig.~\ref{fig:magcap1}(a)]. The magnetic field dependence of N\'{e}el temperature, $T_{N}\left(
H\right) \approx T_{N}\left(0\right) -\lambda H^{2}$, gives rise to a discontinuity of MC at $T_{N}$ and its anomalous behavior below $T_{N}$, is roughly consistent with our data. However, the most prominent feature of the observed anomaly -- the long negative tail for $T > T_{N}$ [see Fig.~\ref{fig:magcap1}(b)] -- cannot be explained in this way.

\begin{figure}[tbp]
\centering
\includegraphics[width=7cm]{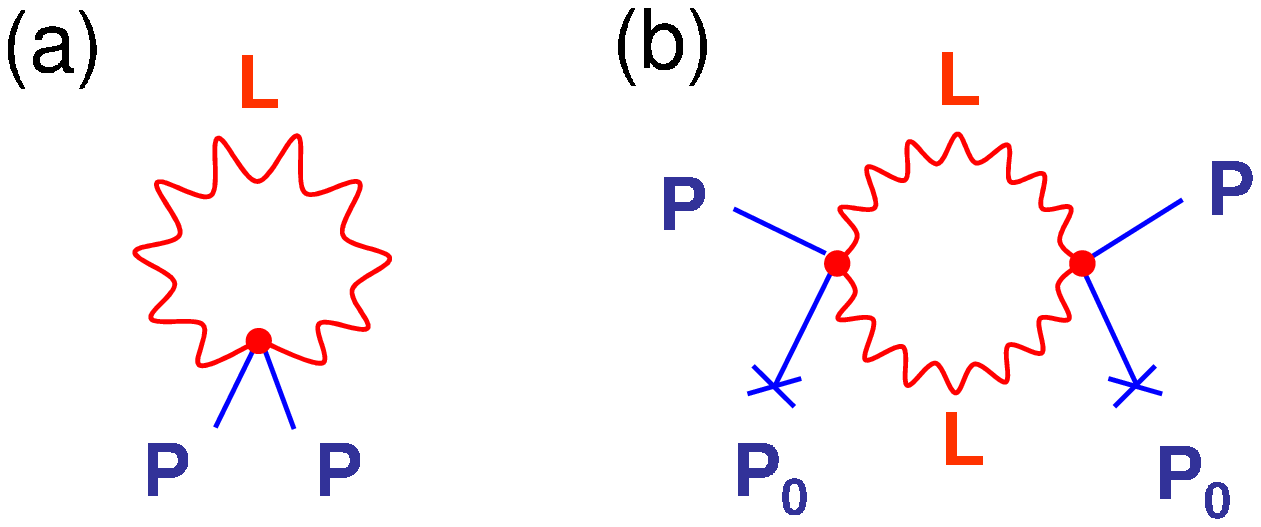}
\caption{(Color online) The self-energy diagrams of the first (a) and second (b) order in the coupling constant $g$, describing contributions of magnetic fluctuations (wavy lines) to dielectric susceptibility.}
\label{fig:Feynman}
\end{figure}

This tail may result from magnetic fluctuations that become critical close to N\'{e}el temperature. The two lowest-order self-energy diagrams describing contributions of magnetic fluctuations to dielectric susceptibility are shown in Fig.~\ref{fig:Feynman}.  The lowest-order term is given by $\delta \chi^{(1)} = - g \chi_{0}^{2} \left(\left \langle L^{2} \right
\rangle - \left \langle L \right \rangle^{2}\right) \propto \tau^{1-\alpha}$, where $\alpha$ is the exponent describing the critical behavior of magnetic specific heat \cite{Katsufuji,LawesPRL2003}. The corresponding singularity in MC $\propto g \lambda \mbox{sign}(\tau)\left| \tau \right|^{- \alpha}$ ($g,\lambda > 0$ for YbMnO$_3$), is positive for $T > T_{N}$, in disagreement with our data.
The second-order fluctuational correction [see Fig.~\ref{fig:Feynman}b],
$
\delta \chi^{(2)} = \frac{2(g \chi_0 P_{0})^{2}}{k_{B}T}\!\int\!\! d^{3}x\!\left[ \left\langle
L^{2}\left( \mathbf{x}\right) L^{2}\left( 0\right) \right\rangle
-\left\langle L^{2}\right\rangle ^{2}\right],
$
results in the MC anomaly $\propto - \lambda \mbox{sign}(\tau)\left| \tau \right|^{-1- \alpha}$, which has the shape similar to that observed experimentally. However, this term is only nonzero in a ferroelectric state with a spontaneous polarization $P_{0} \neq 0$ and, therefore, should be visible in $\varepsilon_{c}$, while we observe an anomaly only in $\varepsilon_{a,b}$.

From this we conclude that the observed anomaly is unrelated to
magnetic fluctuations and originates from a different physics. Below
we show that the shape and scaling behavior of MC can be explained within a mean field theory by the competition
between antiferromagnetic and weakly ferromagnetic state.

Hexagonal manganites show a number of magnetic phases with the 120$^\circ$-angle between  Mn spins in triangular $ab$ layers \cite{FiebigJAP2003}. These phases differ by  orientation of the spins with respect to the crystallographic axes and spins in neighboring layers as well as by the ordering of rare earth spins
\cite{Lonkai2002,Munoz,FiebigPRL2000,FiebigJAP2002}. The magnetic phase diagram of YbMnO$_3$ studied by a variety of different experimental techniques includes the low-field B$_{2}$ phase (magnetic space group P{\underline 6}$_{3}$\underline{c}m) and the high-field A$_{2}$-phase (magnetic space group P6$_{3}$\underline{cm}) \cite{FiebigJAP2003,Yen}. The symmetry of the latter state allows for a net magnetization in the $c$ direction (largely due to the rare earth spins and therefore small near the Mn spin-ordering temperature), which is why the A$_2$ phase is stabilized by $H \| c$.

The competition between the A$_{2}$ and B$_{2}$ phases was discussed in Ref.~\cite{MunawarCurnoe} using the phenomenological free energy expansion in two order parameters:
\begin{eqnarray}
f &=& \sum_{\gamma= \rm {A,B}}\left[\frac{\alpha_{\gamma}}{2}
\left(T - T_{\gamma}^{(0)}+\lambda_{\gamma}H^2\right) L_{\gamma}^2
+\frac{b_{\gamma}}{4} L_{\gamma}^4\right] \nonumber \\
&+& \frac{d}{2}L_{\rm A}^2L_{B}^2 - H \left(\phi L_{\rm A} +
\frac{\phi^{\prime}}{3}L_{\rm A}^3+
\frac{\phi^{\prime\prime}}{2}L_{\rm A}L_{\rm B}^2\right)
\label{eq:freeenergy}
\end{eqnarray}
where $L_{\rm A}(L_{\rm B})$ is the order parameter describing the A$_2$(B$_2$) phase and $H$ is magnetic field along the $c$ axis. The linear coupling of $L_{\rm A}$ to $H$ corresponds to the spontaneous magnetization present in the A$_2$ phase.
\begin{figure}[tbp]
\centering
\includegraphics[width=0.4\textwidth]{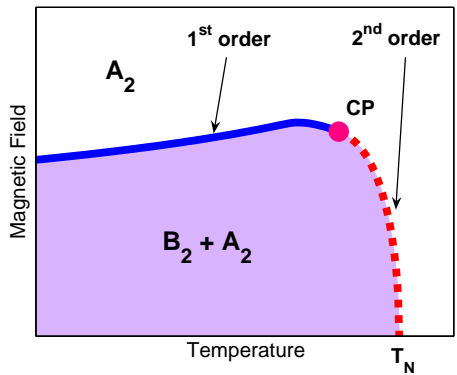}
\caption{(Color online) The magnetic phase diagram of the model Eq.(\ref{eq:freeenergy}) for $T_{\rm B}^{(0)} > T_{\rm A}^{(0)}$. The lilac region is the B$_2$ phase with some admixture of the A$_2$ phase. The critical point (CP) separates the first-order transition (solid) line from the second-order transition (dashed) line.}
\label{fig:phasediagram}
\end{figure}

\begin{figure}[tbp]
\centering
\includegraphics[width=8cm]{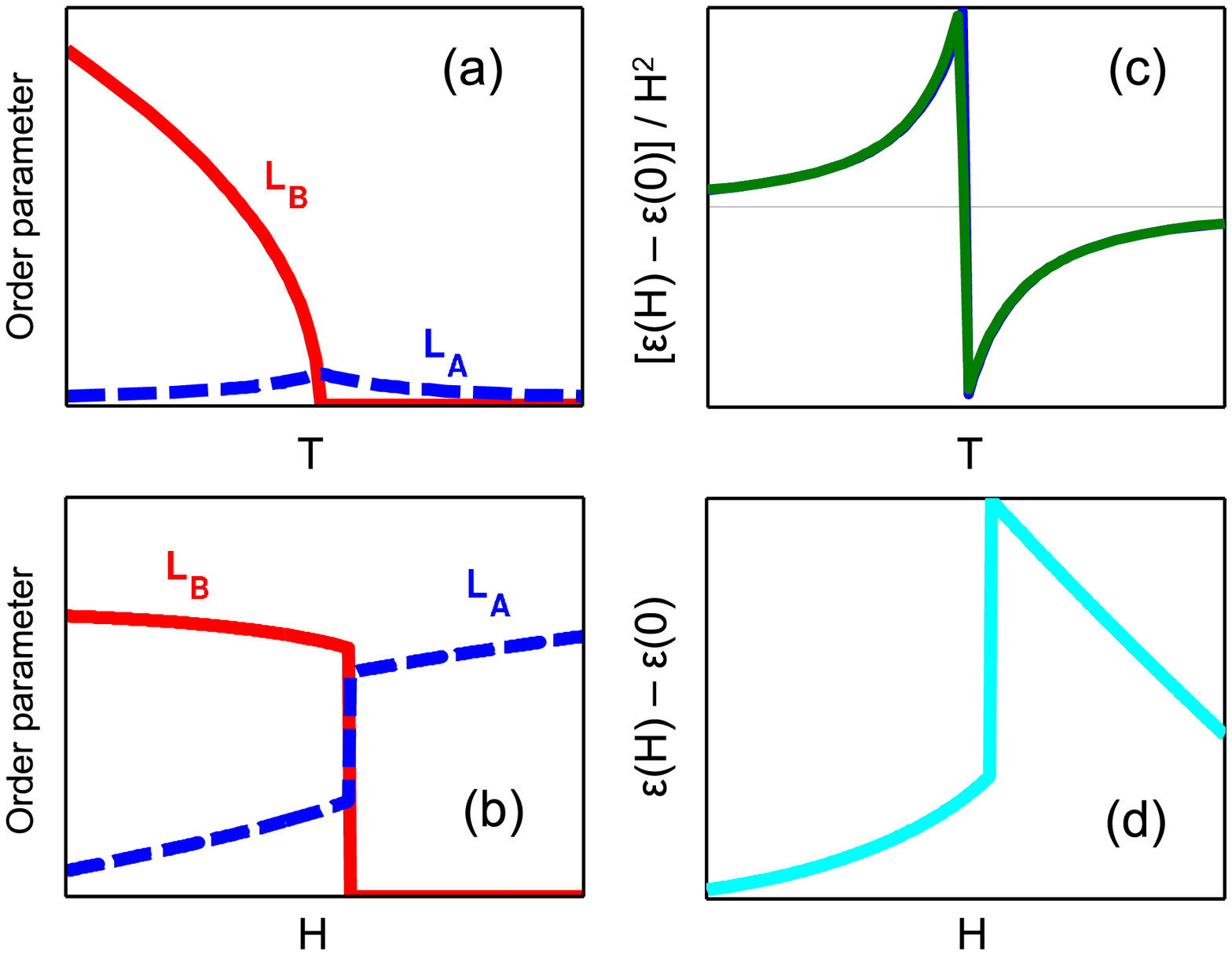}
\caption{(Color online) The temperature dependence of order parameters $L_{A}$ (dashed blue line) and $L_{B}$ (solid red line) near the second-order (a) and first-order (b) transitions. Panel (c) shows the anomaly in rescaled MC at the second-order transition temeprature; the two curves calculated for different values of magnetic field are almost indistinguishable. The MC anomaly at the first-order transition is shown in panel (d).} \label{fig:orderparameter}
\end{figure}

The typical phase diagram for $T_{\rm B}^{(0)} >
T_{\rm A}^{(0)}$ (when the $B_2$ phase is energetically more
favorable than the $A_2$ phase at zero field) is shown in
Fig.~\ref{fig:phasediagram}. Due to the linear coupling between $H$
and $L_{\rm A}$, the latter order parameter is nonzero for an
arbitrarily weak magnetic field, so that for $H \neq 0$ the
transition occurs between the A$_2$ phase and the B$_2$ phase with
some admixture of the A$_2$ phase.

The MC for this model, shown in Fig.~\ref{fig:orderparameter}(c), is calculated by adding to the free energy Eq.(\ref{eq:freeenergy}) the terms describing the coupling of the magnetic order parameters to the in-plane electric polarization and the dielectric response of the nonmagnetic state,
\begin{equation}
\Delta f = \frac{P^2}{2} \left(
\sum_{\gamma=A,B}g_{\gamma}L_{\gamma}^{2} + g_{\rm A}^{\prime}L_{\rm A}H \right) + \frac{P^{2}}{2\chi_{0}} - PE,
\label{eq:mecoupling}
\end{equation}
has the same shape as the one observed in YbMnO$_3$ and, for weak fields, obeys the observed scaling. This behavior can be understood by noting that the main contribution to the magnetic field dependence of the dielectric susceptibility comes from $L_{\rm A}$, which is linearly coupled to $H$. This field-induced order parameter grows as $T$ approaches the $T_{N}$ from above [see Fig.~\ref{fig:orderparameter}(a)], which gives rise to the `high-temperature' negative MC tail, as  $\Delta \chi_{e} \propto - L_{\rm A}^{2}$.      In the weak-field regime $L_{\rm A} \propto H$, so that
$\frac{\chi_{e}(T,H) - \chi_{e}(T,0)}{H^2}$ is approximately field-independent, which explains the observed scaling \cite{signchange}.

\begin{figure}[tbp]
\centering
\includegraphics[width=0.3\textwidth]{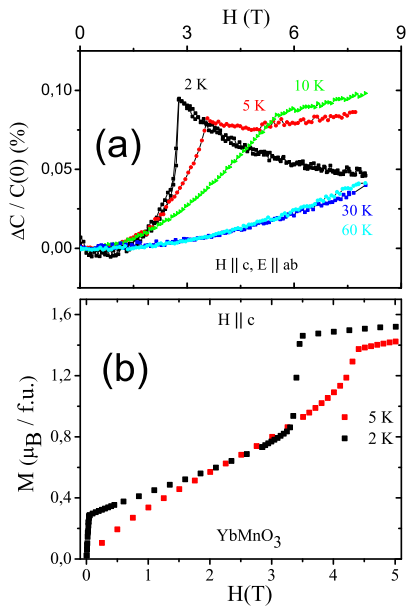}
\caption{(Color online) (a) MC of YbMnO$_{3}$ at $T < T_{N}$ for
$E\| ab$-plane and $H \| c$; (b) Field dependence of magnetization
at 2 and 5K.} \label{fig:lowT}
\end{figure}

As the magnetic field increases, the character of the transition in the two-parameter model changes: in low fields the transition is of second order (red dashed line in Fig.~\ref{fig:phasediagram}), while in high fields and low temperatures it becomes a first-order transition (blue line) [see also Figs.~\ref{fig:orderparameter}(a) and (b)]. The first- and second-order transition lines
are separated by the critical point. This change in the nature of the transition is also clearly seen in the experiments by comparing the field-dependence of MC at low temperatures [see Fig.~\ref{fig:lowT}(a)] to that at high temperatures  [see
Fig.~\ref{fig:magcap1}(a)]. At 2K the MC shows a distinct cusp at the first-order transition, which is well reproduced within our model [see Fig~\ref{fig:orderparameter}(d)].  In
YbMnO$_{3}$ the changes in the order of the transition are made more dramatic by the fact that at low temperatures and high magnetic fields magnetic response is dominated by Yb spins, which below 3.8K order ferrimagnetically \cite{Yen}.  This leads to a strong decrease of the critical magnetic field at low temperatures ($H_{c} \sim 3$T at 2K) and gives rise to the sharp discontinuity in magnetization [see Fig.~\ref{fig:lowT}(b)] \cite{Tomuta}.

Finally, since the shape of the MC anomaly may be affected by the motion of antiferromagnetic domain walls separating different antiferromagnetic phases, we measured the frequency dependence of the dielectric constant. We found no frequency dependence in the interval from 1kHz to 1MHz, suggesting that the contribution of the domain walls in YbMnO$_3$ is negligible.



To summarize, we observed a sharp anomaly in nonlinear
magnetoelectric susceptibility of the hexagonal rare earth manganite YbMnO$_{3}$ at the N\'{e}el temperature. We discussed theoretically possible sources of the anomaly and showed that it results from the competition
between two antiferromagnetically ordered states of YbMnO$_3$, one of which has a small spontaneous magnetic moment. Even though this weakly ferromagnetic phase becomes the ground state only in rather high magnetic fields or at very low temperatures, its admixture to
the non-ferromagnetic phase determines the shape of the
MC anomaly along the whole critical line of magnetic phase transitions. The competition between different magnetic phases, some of which may have weak ferromagnetic moment is very common for frustrated magnets: similar phase diagram were found for hexagonal
HoMnO$_3$, which shows four competing states \cite{FiebigJAP2003} and for Ni$_3$V$_2$O$_8$ \cite{Lawes}. Thus many other systems
should show an anomaly in nonlinear magnetoelectric response, although its shape, which depends on parameters of the Landau free energy, may vary from material to material.

\begin{acknowledgments}
We thank G.R. Blake, G.Nen\'{e}rt and
N. Mufti for useful discussions and J. Baas for technical help. The
work of A.A.N. is supported by the NWO Breedtestrategie Program of
the Material Science Center, RuG and by KNAW, Dutch Royal Academy of
Sciences, through the SPIN program. This work is in part supported
by the Stichting FOM (Fundamental Research on Matter) and in part by
the EU STREP program MaCoMuFi under contract FP6-2004-NMP-TI-4 STRP
033221.
\end{acknowledgments}


\begin{thebibliography}{99}
\bibitem{signchange}
Another contribution to MC comes from an additional increase of $L_{\rm B}$ below $T_N$ due to the suppression of $L_{\rm A}$, which leads to the discontiuity and the sign change of MC at $T_N$. This contribution is also proportional to $H^2$.
\end{thebibliography}
\end{document}